\documentclass[twocolumn,showpacs,showkeys,preprintnumbers]{revtex4}
\usepackage{graphicx}
\usepackage{subfigure}
\usepackage{lscape}
\usepackage{dcolumn}
\usepackage{bm}
\usepackage{color}

\begin{document}

\title{Probing the mass composition of primary cosmic rays from the effect of the geomagnetic field on EAS muons: A simulation study}
\author{Rajat K. Dey}
\email{rkdey2007phy@rediffmail.com}
\affiliation{Department of Physics, University of North Bengal, Siliguri, WB 734 013 India}

\begin{abstract}
The distribution of the muon content of highly inclined Monte Carlo cosmic ray showers is affected by the influence of Earth's geomagnetic field. It is found that the shapes of the positive and negative muon distributions get affected/modified by the influence of the Earth's geomagnetic field. Such a correlation between the earth's geomagnetic activity and the cosmic ray (CR) air shower muons is found sensitive to the primary cosmic ray mass composition.   

\end{abstract}

\pacs{96.50.sd, 95.75.z, 96.50.S} 
\keywords{cosmic-ray, EAS, geomagnetic activity, muons, simulations}

\maketitle

\section{Introduction}
  
The measurement of an extensive air shower (EAS) usually gives electron and muon lateral density distributions (LDD) at a detection level. Timing detectors in the EAS array provides timing information of shower particles. The observed LDD data are usually fitted by means of a lateral density function \emph{e.g.} Nishimura-Kamata-Greisen (NKG) structure function [1] for obtaining the EAS parameters. Timing information provide the estimation of the shower zenith and azimuthal angles. The LDD of cascade particles in an EAS is often assumed to be symmetrical in the shower front plane. However, such an axial symmetry gets affected considerably for showers with $\Theta \geq 50^{\rm o}$ not only due to the intrinsic shower-to-shower fluctuations but significantly by  the earth's geomagnetic field. Such effects distorts the axial symmetry particularly of the distribution of muons in an EAS. Inclined showers manifests significant asymmetries due to such a geomagnetic effect.

In data analysis, one needs to project the detector signals known at the ground plane onto the shower front plane. During the implementation of the above projection, contribution arising from the shower evolution of the late regions is usually being ignored. If such a contribution is taken into account the circular symmetry in the LDD data estimated from ground detectors is broken for inclined showers. The LDD data rather manifests polar asymmetry in the shower plane. This altogether refers to the geometrical and attenuation effects to polar asymmetries of an EAS [2]. A correlation of Earth's geomagnetic activity with the magnitude of asymmetry of the LDD of muons can be known if the geometric and attenuation effects are removed in the EAS data analysis. This work looks for extracting some imprints of the geomagnetic field (GMF) on LDD of muons of EASs. The asymmetric distribution caused by the GMF can then be quantified by an observable, called the transverse muon barycenter separation (TMBS). The TMBS is defined as the linear distance between the center of gravities of positive and negative muons in the shower plane.

\begin{figure}
\includegraphics[scale=0.25]{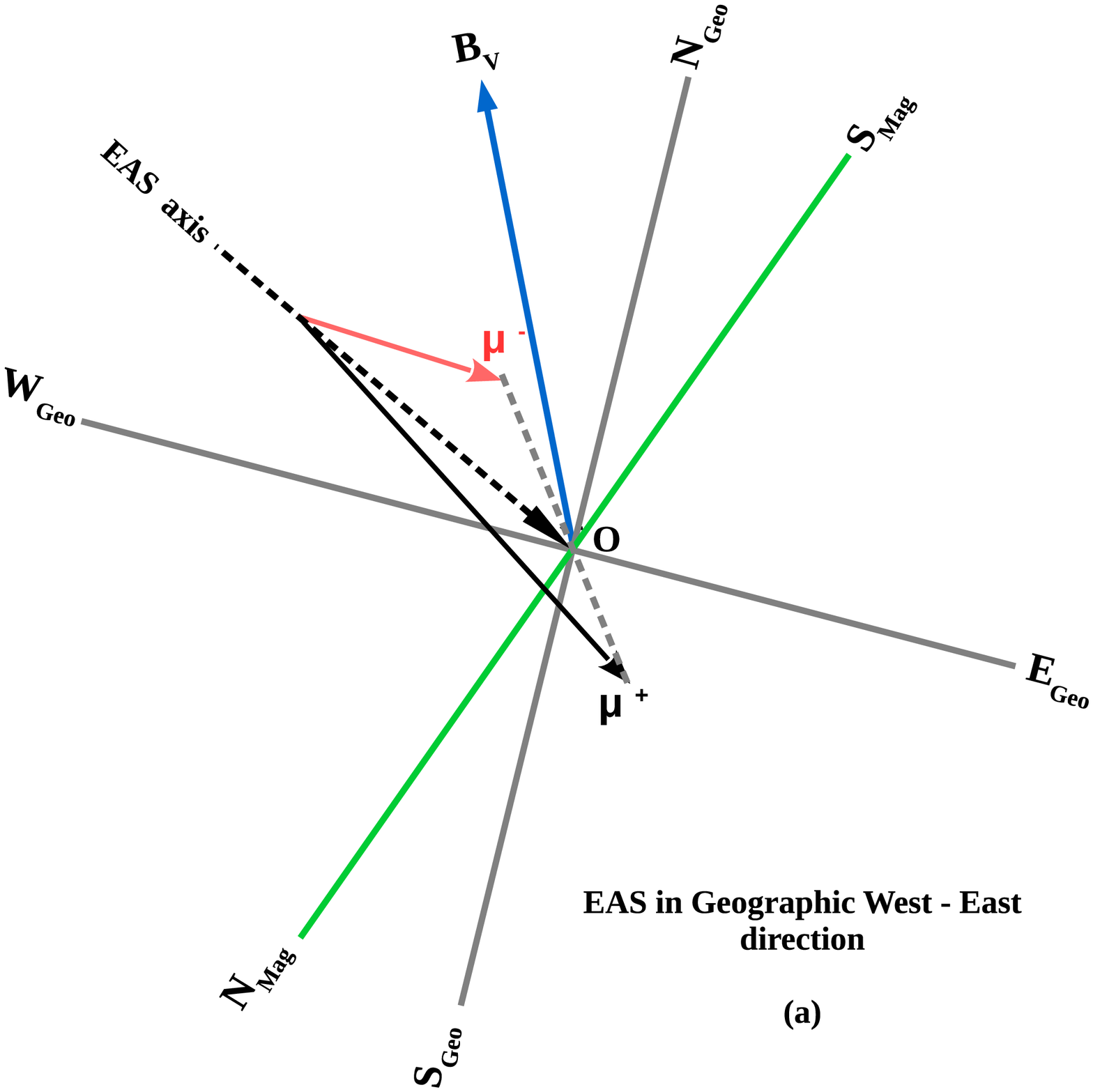} 
\includegraphics[scale=0.25]{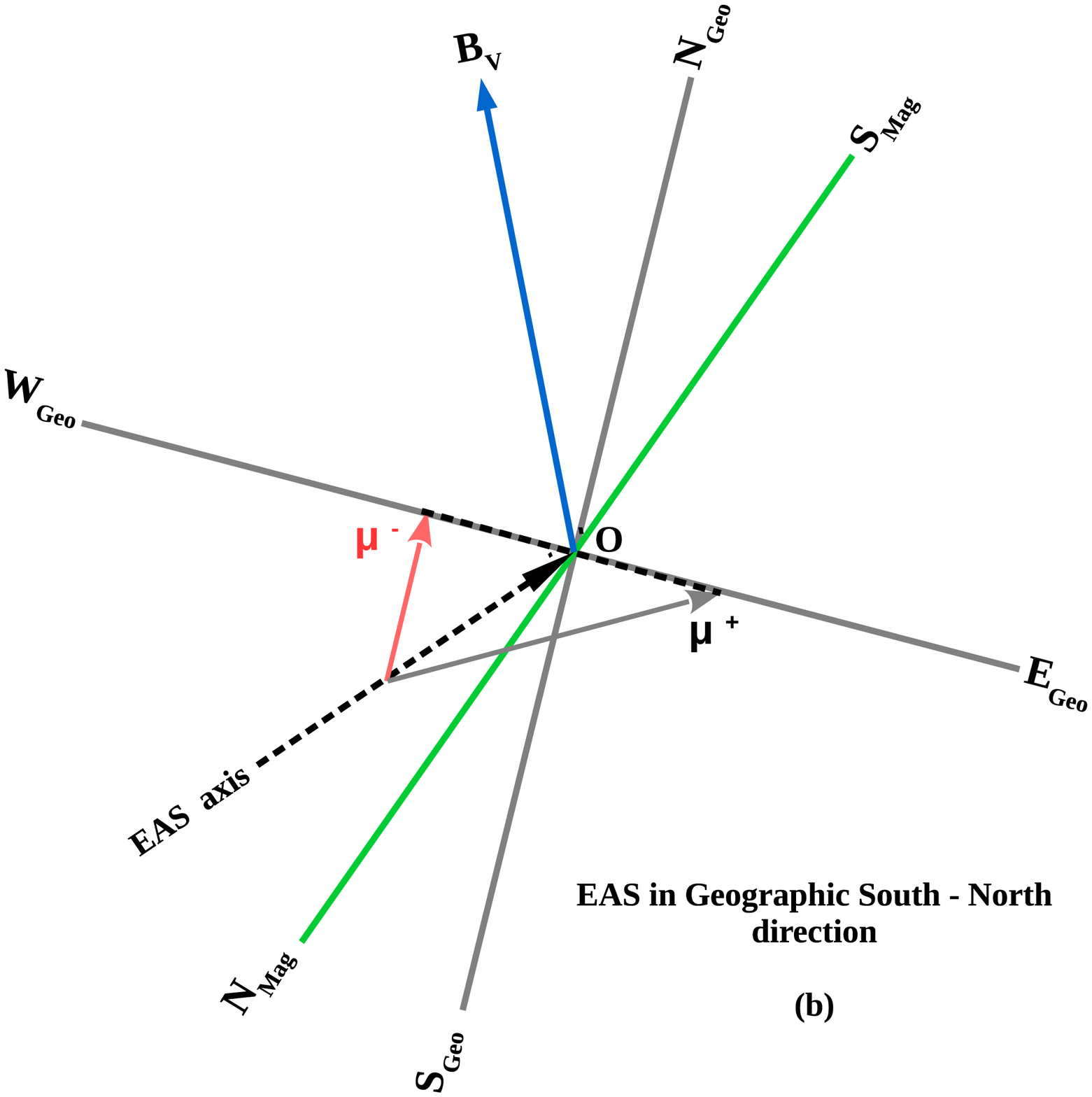}
\caption{The motion of $\mu^{+}$ and $\mu^{-}$ originated from a parent particle in shower experiencing the GMF at a place with east magnetic declination. The pair of lines represent the geographic and magnetic meridians respectively.}
\end{figure} 

In the shower cascade, electrons have shorter radiation lengths suffering strong attenuation as well bremsstrahlung effects in the atmosphere. These electrons thereby will rapidly change the directions of their momenta relative to the GMF. These processes cause wider lateral spread of electrons in the EAS, and hence the effect of GMF on them is insignificant. However, muons travel large distances in the air encountering negligible scattering and hence come under the influence of GMF significantly [3]. Fig. 1 showing the deflections of a $\mu^{+}$ and a $\mu^{-}$ by a GMF generated usually from a charged pion in an EAS arriving from two different directions. The Fig. 1a shows the deflections of $\mu^{+}$ and $\mu^{-}$ in opposite directions generated from a cosmic ray (CR) particle coming from the geographic west and advances to the east direction. But, in the Fig. 1b where the EAS comes from the geographic south and advances into the north direction. 

In this paper the asymmetry of muons arising from the influence of the GMF on them is investigated via Monte Carlo (MC) simulation, using the air shower simulation code {\emph{CORSIKA}} [4]. The work mainly discusses the impact of Earth's magnetic field components on the TMBS parameter and its maximum value i.e. MTMBS at the KASCADE location. Finally, sensitivity of the MTMBS parameter mainly to CR mass composition is demonstrated.

The paper is organized as follows. The MC code {\emph{CORSIKA}} adopted here is discussed in section II. The main data analysis technique for obtaining the crucial observables is discussed in section III. The results are presented and briefly discussed in section IV. Section V presents the summary of our conclusions.

\section{Monte Carlo simulations}
The \emph{CORSIKA} [4] is a MC simulation code that simulates the evolution of a CR shower in the atmosphere. The \emph{CORSIKA} showers are generated with the combination where the hadronic model UrQMD [5] acts at low energies ($\rm{E_{h}} < 80 {\rm{GeV/n}}$) and the EPOS $1.99$ model [6] as a high energy hadronic model. We consider the U.S. standard atmospheric model as the shower development medium in the code [4]. The electromagnetic (EM) interactions was addressed by the EGS$4$ program library [7]. 

The KASCADE air shower array [8-9] site is chosen as the level for the generation of simulated showers. The simulated showers have been generated at a fixed 1 PeV primary energy with spectral index $\gamma \sim 2.7$. More showers are generated in the primary energy regions: $1-3$, $8-12$ and $98-102$ PeV in the simulation for our present purpose. The kinetic energy cut-offs are set as $0.003$ GeV for electrons and $0.3$ GeV for muons in the CORSIKA steering file. Only proton (p) and iron (Fe) shower are simulated for the purpose. Inclined showers with $\Theta \geq 50^{\rm o}$ are generated to receive more influence by the GMF on muons in particular. The azimuthal angle ($\Phi$)varies from $0$ to $2\pi$ in the simulation runs. To observe the effect of the GMF on muons, a large number of showers are also generated by keeping Earth's magnetic field in switch-off mode in the simulation. 

\section{Data analysis procedure and selection cuts}

To obtain the imprint only due to the GMF effect in the azimuthal distribution of high energy muons, the following data analysis procedure has been undertaken.

For highly zenith angle showers, the information like the $\rho_{\mu}$ or $N_{{\mu}^{\pm}}$ get overestimated in the advanced region while underestimated in the late part of of a shower. It tells that muons in the late region do undergo higher attenuation than those arriving in the advanced region in an EAS. This accounts the attenuation effect to the overall polar asymmetry of the distribution of muons and is known as the attenuation effect on an EAS. The polar asymmetries are to be modified while the information of muons is being transformed from the array plane to the shower plane, and is termed as the geometric effect.

\begin{figure}
\includegraphics[scale=0.35]{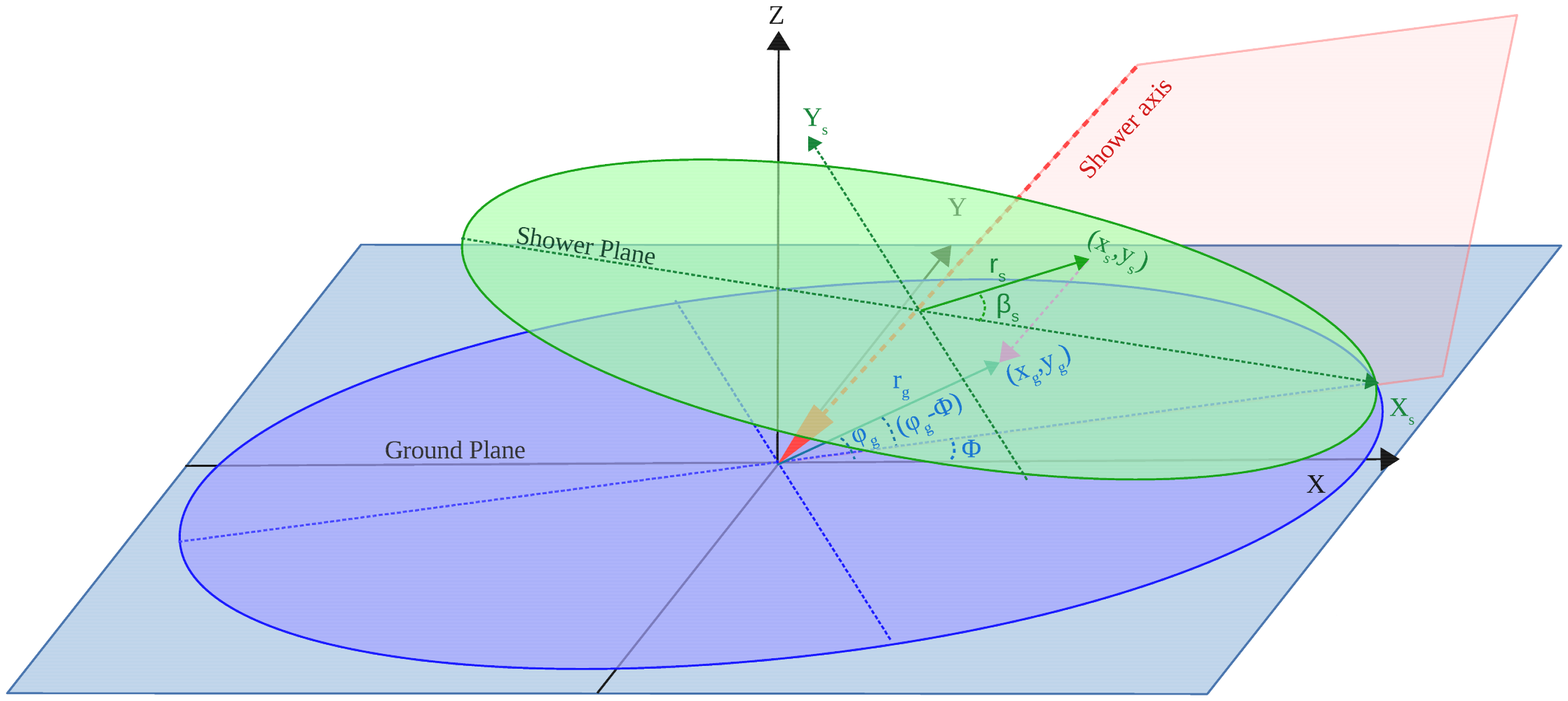}
\caption{Geometry of the array plane and shower front plane used for the geometric correction in an inclined shower. $\Theta$ and $\Phi$ are shower zenith and azimuth angles.}
\end{figure}

\begin{figure}
\includegraphics[scale=0.3]{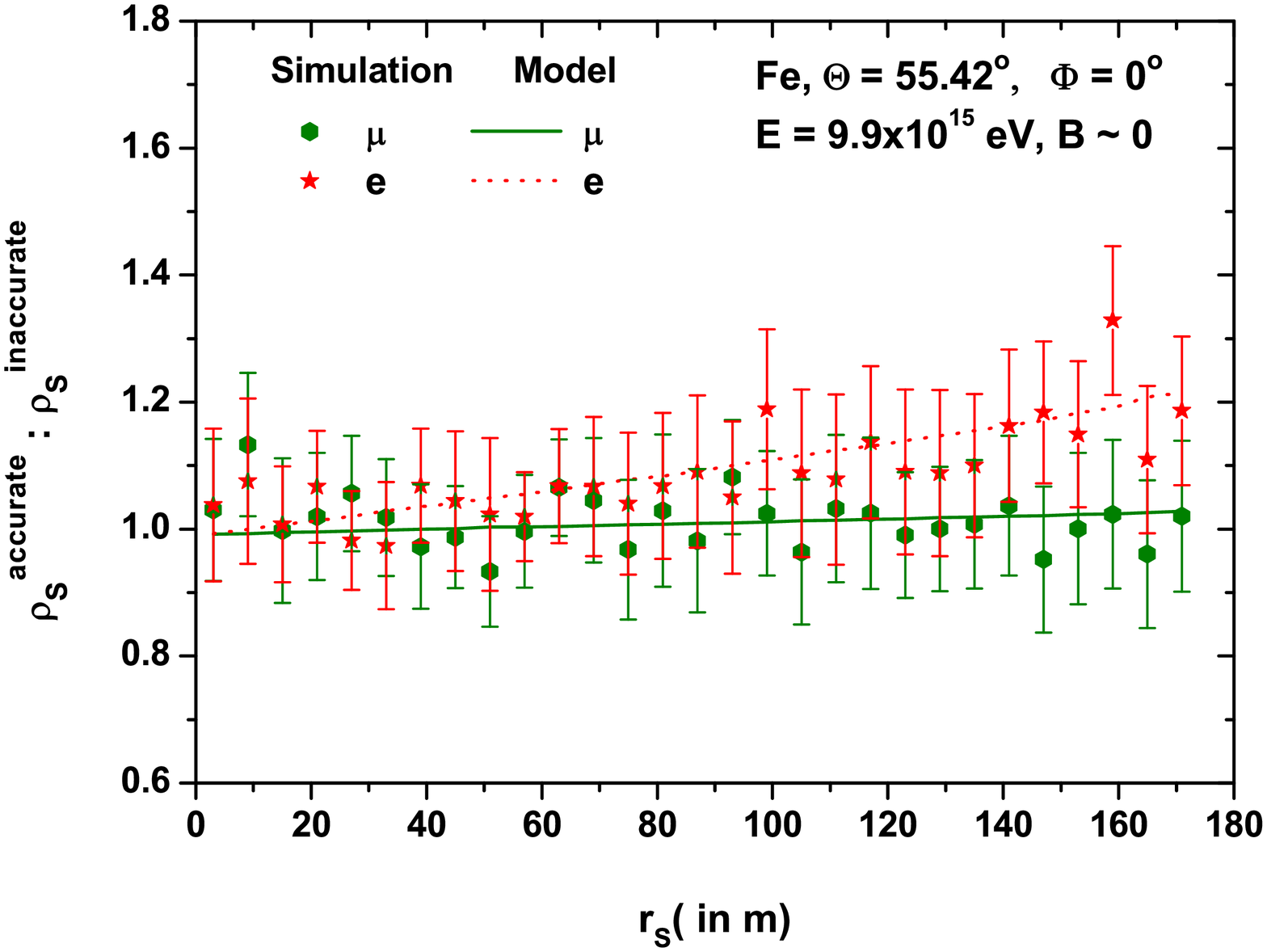}
\caption{The ratios of the accurate to the inaccurate densities for showers coming from the magnetic north with core distance in the shower front plane. The attenuation model gives the ratio as, $\rm{e}^{-({\eta x_{g}sin\Theta}}) \approx (1-{\eta \rm{x}_{g}sin\Theta})$ for the delayed part of the shower with $\rm{x_{g}}$ is negative. Two lines describing the ratios from the model.}
\end{figure} 

Projecting first the position of hit of each muon from the ground plane onto the shower plane in our analysis. According to the Fig. 2, the transformation of a point of hit by a muon in the ground plane with coordinates; ($r_{\rm g}$,$\phi_{\rm g}$ or $\rm{x}_{\rm g}$, $\rm{y}_{\rm g}$) onto the shower plane with the new set as; ($r_{\rm s}$,$\beta_{\rm s}$ or $\rm{x}_{\rm s}$, $\rm{y}_{\rm s}$), is described. The required relations are as follows,
  
\begin{equation}
r_{\rm{s}} = r_{\rm{g}} \sqrt{1-sin^{2}\Theta cos^{2}(\phi_{\rm{g}}-\Phi)}
\end{equation}

or, equivalently;
\begin{equation}
x_{\rm{s}} = x_{\rm{g}}cos\Theta = r_{\rm{g}}cos(\phi_{\rm{g}}-\Phi)cos\Theta
\end{equation} 
\begin{equation} 
y_{\rm{s}} = y_{\rm{g}} = r_{\rm{g}}sin(\phi_{\rm{g}}-\Phi)
\end{equation}

The muon density in the shower plane ($\rho_{\rm s}(r_{\rm s},\beta_{\rm s}$)) can be known from the muon density estimated in the ground plane ($\rho_{\rm g}(r_{\rm g},\phi_{\rm g}$)) by a  transformation, and is given by,
         
\begin{equation}
\rho_{\rm{s}}^{\rm{incorr.}}(r_{\rm{s}},\beta_{\rm{s}}) = \frac{\rho_{\rm{g}}(r_{\rm{g}},\phi_{\rm{g}})}{cos{\Theta}}.
\end{equation} 

The $\rho_{\rm s}^{\rm{inaccur.}}(r_{\rm s},\beta_{\rm s}$) obtained from the above, is inaccurate as because they do not at all involve the attenuation of muons in the region between the two planes. An accurate measure for the muon density ($\rho_{\rm s}^{\rm{accur.}}(r_{\rm s},\beta_{\rm s}$)) in the shower plane [10], taking account the geometrical projection of $\rho_{\rm g}(r_{\rm g},\phi_{\rm g}$) onto the shower plane including the attenuation effect, is done through
 
\begin{equation}
\rho_{\rm{s}}^{\rm{inaccur.}}(r_{\rm{s}},\beta_{\rm{s}}) = \frac{\rho_{\rm{g}}(r_{\rm{g}},\phi_{\rm{g}})}{cos{\Theta}}~e^{{\pm}({\eta x_{g}sin\Theta})},
\end{equation} 
  
where $\eta$ = $\frac{\Delta{X}}{\lambda}$ accounts the attenuation length in $\rm m^{-1}$. The attenuation length $\lambda$ for electrons at the KASCADE level is about 190 g cm$^{-2}$. For the muonic component, $\lambda$ takes a value close to 900 g cm$^{-2}$. Near the surface of the Earth the increase $\Delta{X}$ is approximately equal to 0.15 g cm$^{-2}$ for every meter travel. Hence $\eta$ takes $\sim \frac{0.15}{190}$ and $\sim \frac{0.15}{900}$ m$^{-1}$ respectively for $\rm e^{\pm}$ and ${\mu}^{\pm}$ at the KASCADE level [11,12]. For a cylindrical EAS profile, we have to substitute $\pm{x_{\rm g} sin\Theta}$ ($-$ and $+$ signs accounting the attenuation of the late and advanced parts of the EAS) for the additional path length between the planes. Equations 4 and 5 are working for the showers coming from the north direction i.e. $\Phi = 0^{\rm o}$ in \emph{CORSIKA} plane. For an arbitrary direction ($\Phi$), one should use $\phi_{\rm g}-{\Phi}$ for $\phi_{\rm g}$ in those equations. We will now show the ratio between the accurate and inaccurate muon densities against the radial distance (${r_{\rm s}}$) for simulated muon densities when ${\rm B}$ is $\sim 10^{-5}\times {\rm B}_{\rm{KAS.}} \sim 0$. The geometry of Fig. 2 giving the following  

 \begin{equation}
{r_{\rm s}} = (x_{\rm g}^{2}cos^{2}\Theta + y_{\rm g}^{2})^{\frac{1}{2}}.
\end{equation}

 At ${\rm B} \sim 0$, the ratio between the accurate and inaccurate muon or electron densities in the delayed part of the shower plane are shown in the Fig. 3 with ${r_{\rm s}}$. The scattered points represent simulated data while the lines are obtained directly from the equations 4 and 5 [10].  It reveals that the attenuation of muons in the region between the two planes can be made to zero judiciously in the data analysis.

The TMBS and MTMBS parameters get affected by the energy of incoming muons. A best compromise among the muon detector size, muon energy thresholds, and the $N_{\mu}^{\rm{tr.}}$ must be established. The parameter $N_{\mu}^{\rm{tr.}}$ accounts the total number of muons in a region between $60$ m and $90$ m core distances with $10^{\rm o}$ or $15^{\rm o}$ polar angle bin made by two diagonal lines passing through the EAS core [13]. The noticeable effects of the GMF are emphasized in the case of highly inclined showers with high momentum muons ($10^2 - 10^3$ GeV/c) [13-15]. 

\section{Results and discussions}
\subsection{Asymmetry of high energy muons in EAS}

We have estimated total number of $\mu^{+}$ and $\mu^{-}$ over a region between the arc of radii $60$ m and $90$ m, with a central angle of amount $\sim 15^{\rm o}$ on the shower plane when GMF is switched-on (taking $\rm{p}_{\mu} = 10^2 - 10^3$ GeV/c). The Fig. 4a shows polar asymmetries of $\mu^{+}$ and $\mu^{-}$ for $\langle \Theta\rangle = 65.44^{\rm o}$ and the selected $\langle \Phi\rangle = 52.5^{\rm o}$ at KASCADE level. We have also performed the same study to obtain the Fig. 4b but for showers having $\langle \Phi\rangle = 245^{\rm o}$. It is noticed from these figures that for $\langle \Phi\rangle = 52.5^{\rm o}$, the modulations of muon species are seen around $\beta_{\rm s} \sim 165^{\rm o}$ and $\sim 345^{\rm o}$. But, for $\langle \Phi\rangle = 245^{\rm o}$, around $\beta_{\rm s} \sim 132^{\rm o}$ and $\sim 312^{\rm o}$ indicate the prominent modulation regions.
\begin{figure}
\includegraphics[scale=0.2]{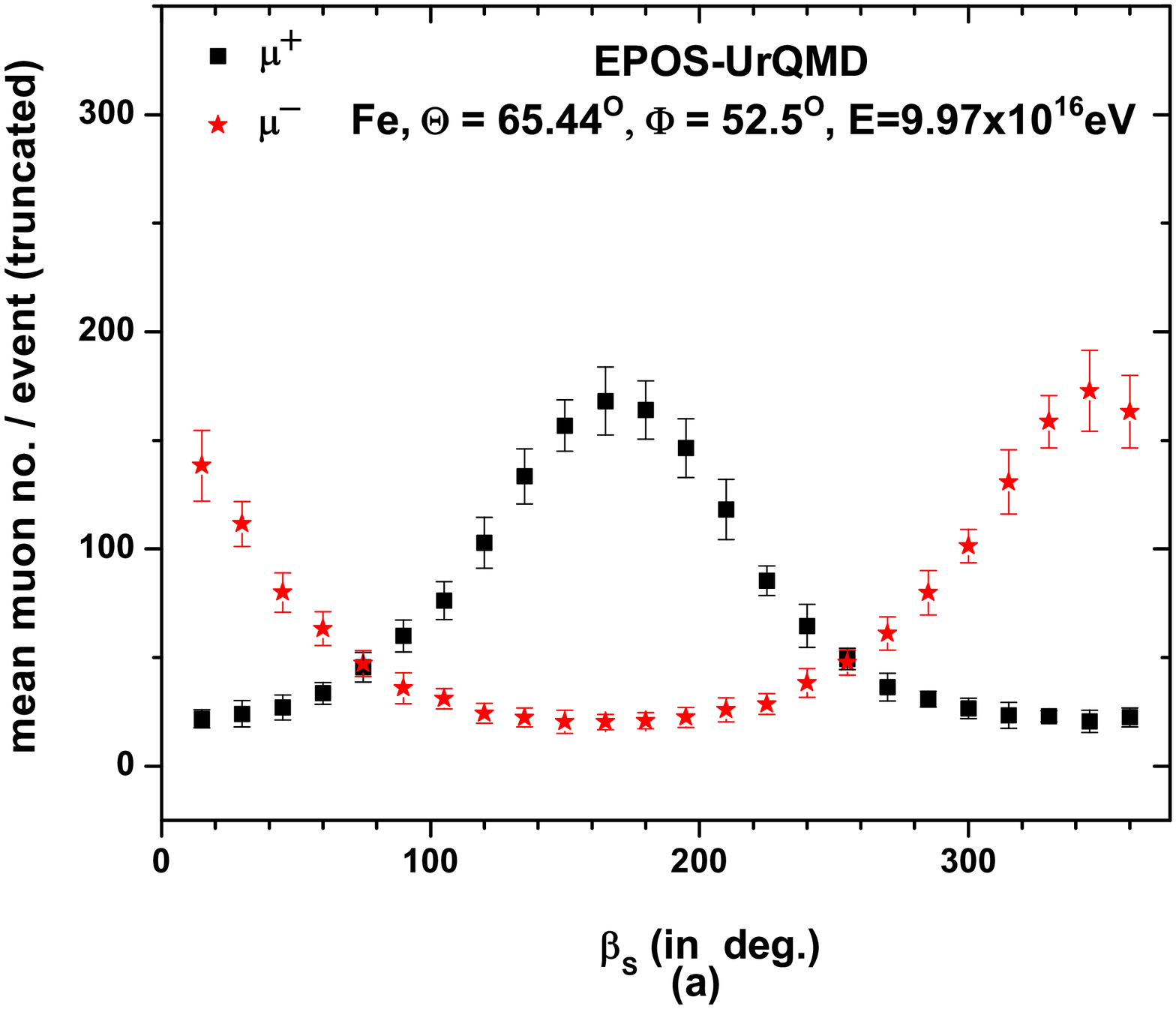} 
\includegraphics[scale=0.2]{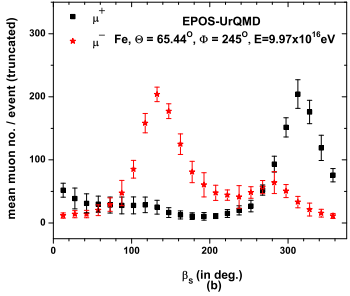} 
\caption{In Fig. a and Fig. b, we have used our two selected $\Phi$ ranges but keeping the mean $\Theta$ at $65.44^{\rm o}$ for the mean polar variations of $\mu^{+}$ and $\mu^{-}$ for iron primary. Here, the x-label represents the polar angle; $\beta_{\rm s}$. Data from the shower plane are used only.}
\end{figure}
 \begin{figure}
\includegraphics[scale=0.2]{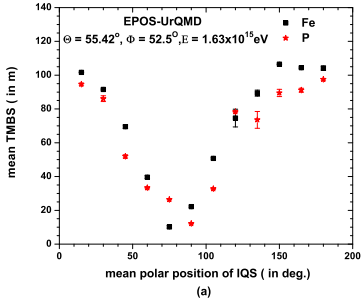} 
\includegraphics[scale=0.2]{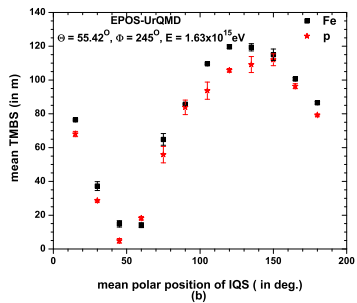} 
\caption{Polar variation of the mean TMBS for p and Fe showers arriving from two mean arbitrary directions.}
\end{figure} 
The TMBS parameter, following the IQS position with highest concentrations of $\mu^{+}$ and $\mu^{-}$ in a diagonally opposite regions, has been estimated finally. This is called the maximum TMBS (i.e. MTMBS) parameter, and such a parameter will be used extensively to correlate Earth's geomagnetic activity with high energy muons.
\begin{figure}
\includegraphics[scale=0.2]{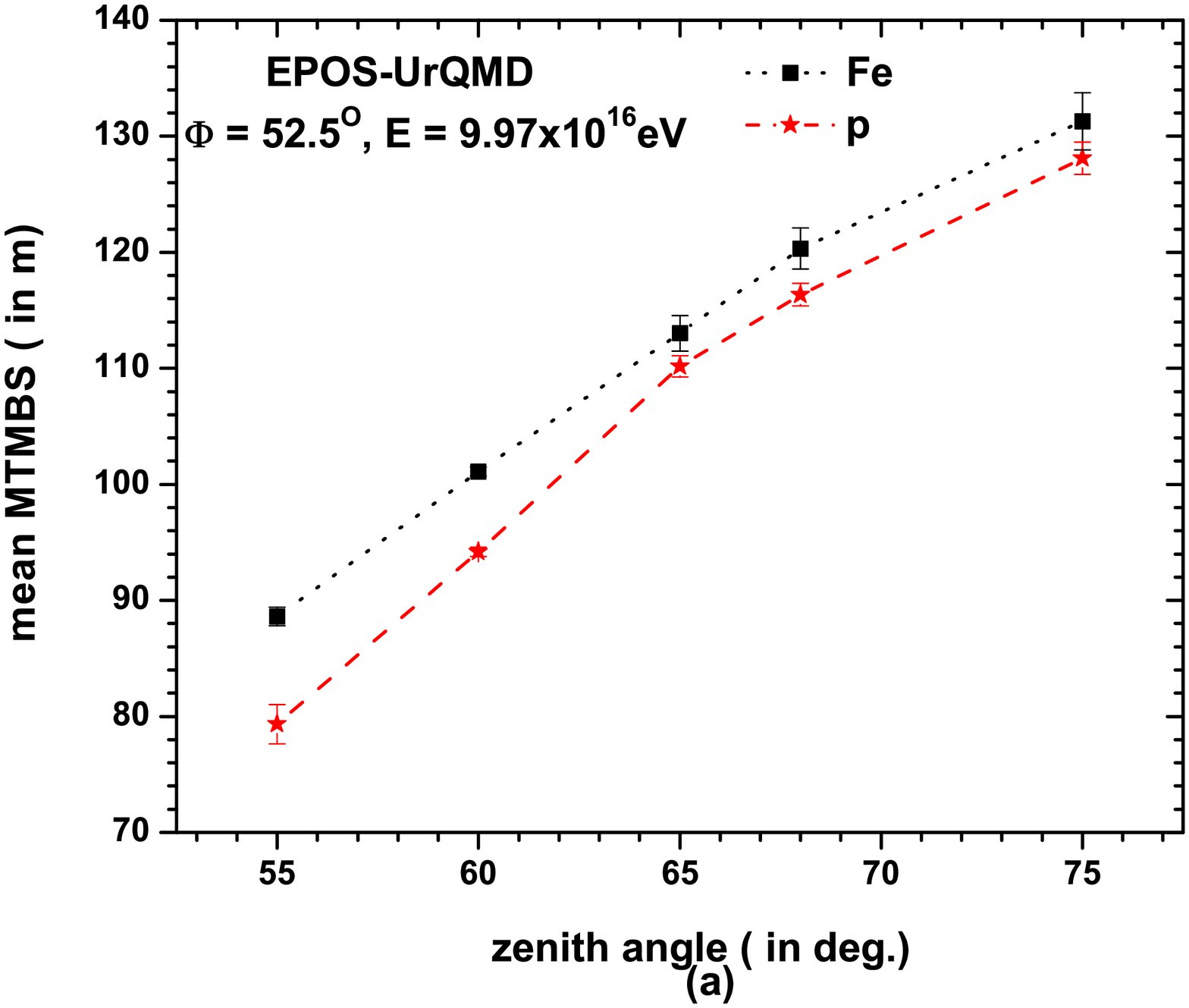}
\includegraphics[scale=0.2]{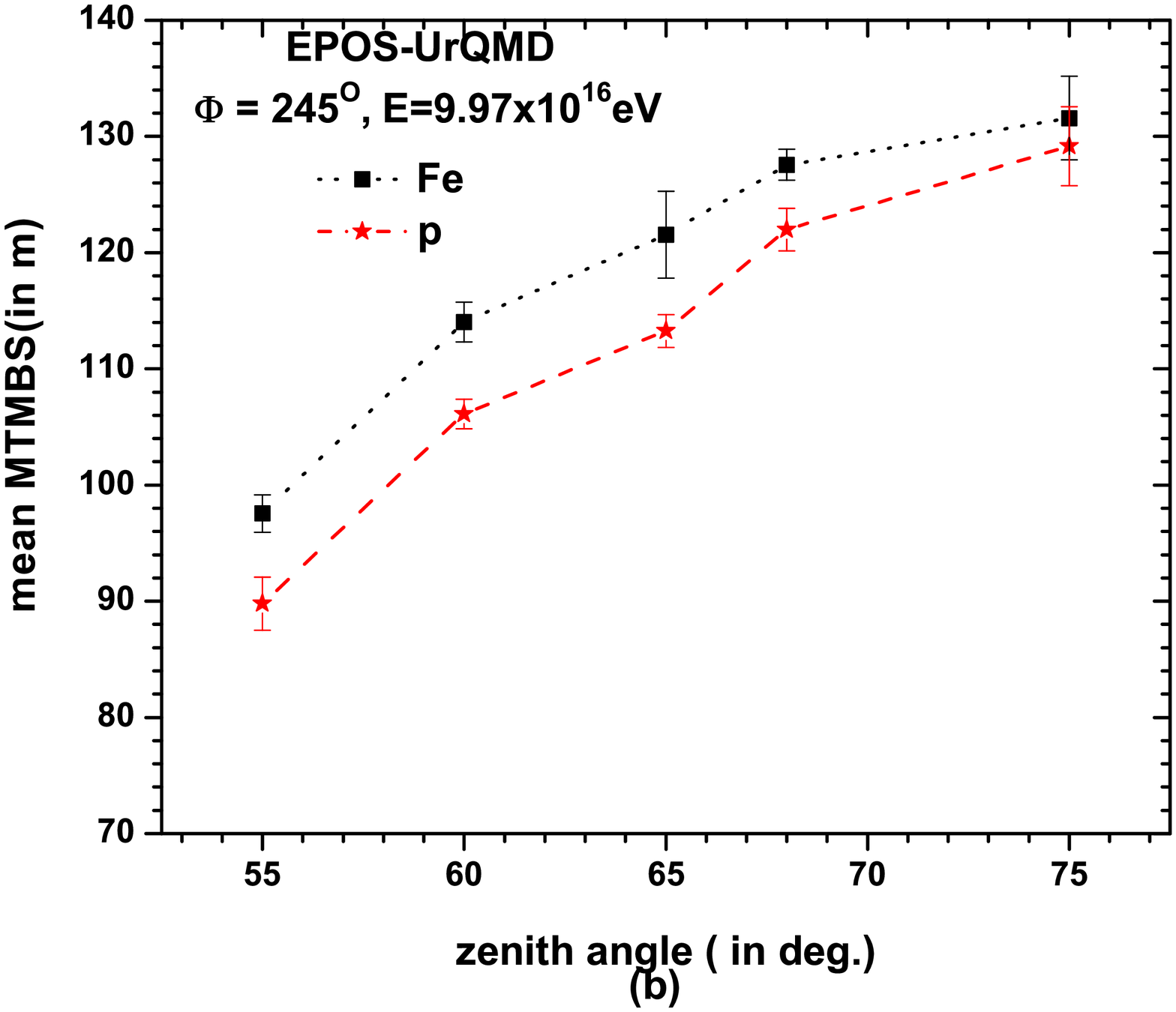}
\includegraphics[scale=0.2]{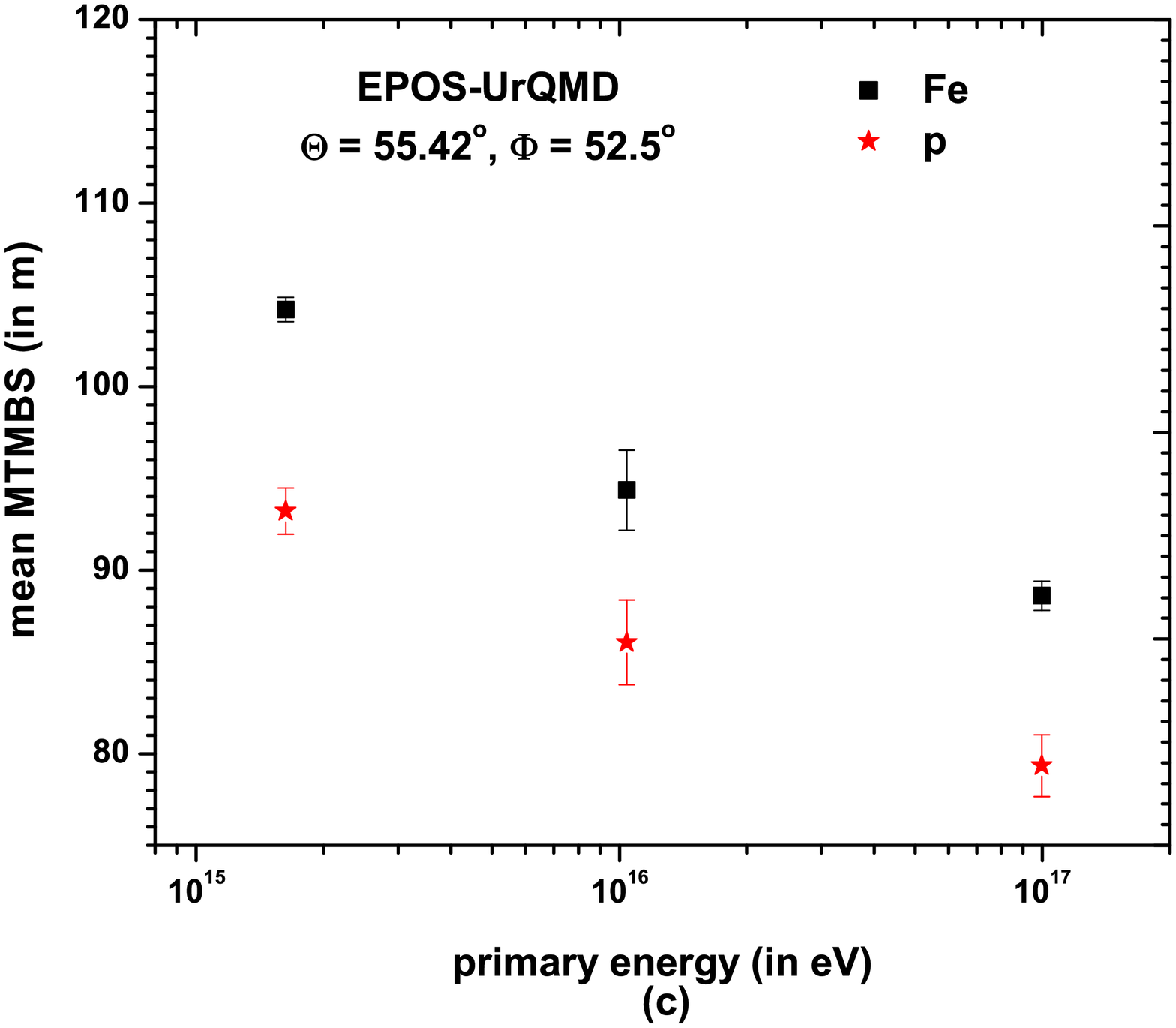}
\includegraphics[scale=0.22]{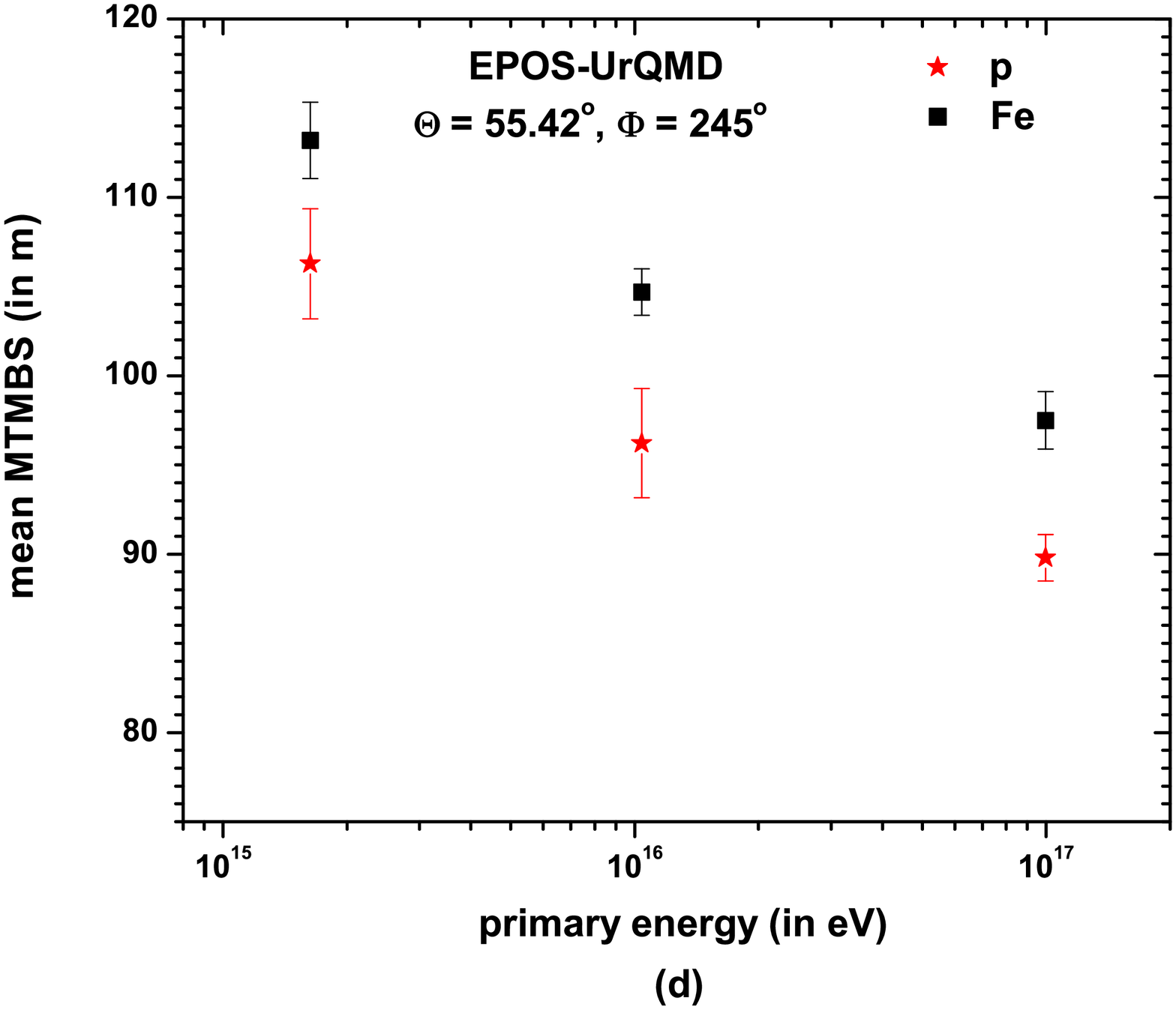}   
\caption{Dependence of the mean MTMBS parameter on the zenith angle and the primary energy of CR showers.}
\end{figure} 
We are now in a position to estimate the coordinates of barycenters of $\mu^{+}$ and $\mu^{-}$ in the shower plane for each shower. The linear distance between the barycenters of $\mu^{+}$ and $\mu^{-}$ from two opposed regions in a hypothetical interior quadrant sector (IQS) is then estimated. This distance is just equal to our desired TMBS parameter in the work. To estimate the parameter at multi-polar positions, an operation is introduced that executes a rotation either clockwise or anti-clockwise sense of the IQS for estimating the $\mu^{+}$ and $\mu^{-}$ positions. An IQS represents a region in the interior between two circles enclosed by a pair of arcs on opposite sides and a pair of diagonally aligned straight lines passing through the EAS core making a central angle of $\sim 10^{o}/15^{\rm o}$.

The TMBS is expected to vary with the rotation of IQS. To show that we have used p and Fe initiated EASs arriving from average direction, $\langle \Phi\rangle = 52.5^{\rm o}$ with the zenith angle range $53^{\rm o} - 58^{\rm o}$. In Fig. 5a such a variation is depicted. Similarly for showers coming from an azimuthal direction $\langle \Phi\rangle = 245^{\rm o}$, our results are shown in the figure 5b. Through the Fig. 6, the CR mass sensitivity of the MTMBS parameter is also demonstrated by studying the dependencies of the MTMBS parameter with $\Theta$ and $\rm E$. 

\section{Summary}
In this work we have described an analysis of the asymmetric muon distributions in highly inclined showers under the effect of Earth's GMF. The present analysis has explored the expected asymmetry of muons as a function of the observable, TMBS/MTMBS. The measurement of the muon charge sign is the primary concern in any concept of designing a possible experimental set up for implementing the proposed method in the work. It has been understood that the the primary CR mass composition at least in the energy region $1 - 100$ PeV can be known by applying the present technique.

\end{document}